\documentclass[conference, 10pt]{IEEEtran}
\IEEEoverridecommandlockouts 
\usepackage[]{graphicx}
\usepackage{caption}
\usepackage{subfig} 
\usepackage{amsmath}
\usepackage{amssymb}
\usepackage{epsfig}
\usepackage{cite}
\usepackage{color}
\usepackage{balance}
\usepackage{amsmath}
\usepackage{amsfonts} 
\usepackage{graphicx}
\usepackage{pdfpages}
\usepackage[T1]{fontenc} 
\usepackage{array}
\usepackage{url}

\usepackage{color}
\usepackage{wrapfig}
\usepackage{lipsum}
\usepackage[hidelinks]{hyperref}
\usepackage{multirow}
\usepackage{tabularx}
\usepackage{tikz}
\usepackage{nth} 
\usepackage{algpseudocode} 
\usepackage{algorithm,algpseudocode}
\begin{document}
\title{Cloud Radiative Effect Study Using Sky Camera}
\author{\IEEEauthorblockN{Soumyabrata Dev\IEEEauthorrefmark{1},
Shilpa Manandhar\IEEEauthorrefmark{1},
Feng Yuan\IEEEauthorrefmark{1},
Yee Hui Lee\IEEEauthorrefmark{1} and
Stefan Winkler\IEEEauthorrefmark{2}}
\IEEEauthorblockA{\IEEEauthorrefmark{1}School of Electrical and Electronic Engineering, Nanyang Technological University (NTU), Singapore}
\IEEEauthorblockA{\IEEEauthorrefmark{2}Advanced Digital Sciences Center (ADSC), University of Illinois at Urbana-Champaign, Singapore}
\thanks{This research is funded by the Defence Science and Technology Agency (DSTA), Singapore.}\thanks{Send correspondence to S. Winkler, E-mail: Stefan.Winkler@adsc.com.sg.}
\vspace{-0.6cm}
}

\maketitle

\begin{abstract}
The analysis of clouds in the earth's atmosphere is important for a variety of applications, viz.\ weather reporting, climate forecasting, and solar energy generation. In this paper, we focus our attention on the impact of cloud on the total solar irradiance reaching the earth's surface. We use weather station to record the total solar irradiance. Moreover, we employ collocated ground-based sky camera to automatically compute the instantaneous cloud coverage. We analyze the relationship between measured solar irradiance and computed cloud coverage value, and conclude that higher cloud coverage greatly impacts the total solar irradiance. Such studies will immensely help in solar energy generation and forecasting. 
\end{abstract}

\IEEEpeerreviewmaketitle

\section{Introduction}
Cloud analysis is traditionally performed primarily via satellite images. However, because of low temporal and spatial resolution of these images, and localized weather phenomenon, such cloud analysis are now performed using ground-based sky cameras and human observations. Clouds affect the incoming solar irradiance, and it is one of the important factors in the intermittency and variability of obtained solar radiation. Luo et al.\ in \cite{Luo2010} used the solar radiation observations to estimate total cloud cover obtained from human observations. Recently, Nikitidou et al.\ also studied the effect of clouds on solar radiation using an all-sky cameras~\cite{Nikitidou2017}. In this paper, we investigate the impact of cloud coverage on the variation of received solar radiation obtained on the earth's surface. This will greatly help in solar energy generation and photovoltaics planning and installation.

\section{Clear Sky Model and Methodology}
\subsection{Clear Sky Model}
\label{sec:CSM}
The solar radiation recorded on the earth's surface on a clear-sky day follows a cosine response with respect to the time of the day. In the literature, there are several clear-sky models that estimates the total solar irradiance at a particular location. For Singapore, the best performing model is proposed by Yang et al.\ \cite{dazhi2012estimation}. It estimates the clear-sky Global Horizontal Irradiance (GHI), measured in W/$\mbox{m}^2$ as: 

\begin{align}
\label{eq:GHI-model}
G_c = 0.8277E_{0}I_{sc}(\cos\theta_z)^{1.3644}e^{-0.0013\times(90-\theta_z)}.
\end{align}

In Eq.~\ref{eq:GHI-model}, $I_{sc}$ is a solar irradiance constant ($1366.1$ W/$\mbox{m}^2$), $\theta_z$ is the solar zenith angle (in degrees). The term $E_{0}$ is a correction factor that is given by:

\begin{equation*}
\begin{aligned}
\label{eq:E0value}
E_0 = 1.00011 + 0.034221\cos(\Gamma) + 0.001280\sin(\Gamma) + \\0.000719\cos(2\Gamma) + 0.000077\sin(2\Gamma),
\end{aligned}
\end{equation*}

,where $\Gamma = 2\pi(d_n-1)/365$ is the day angle (in radians), and $d_n$ is the day number in a year.

\subsection{Cloud Radiative Effect}
\label{sec:CREffect}
\begin{figure}[htb]
\begin{center}
\noindent
\includegraphics[width=0.47\textwidth]{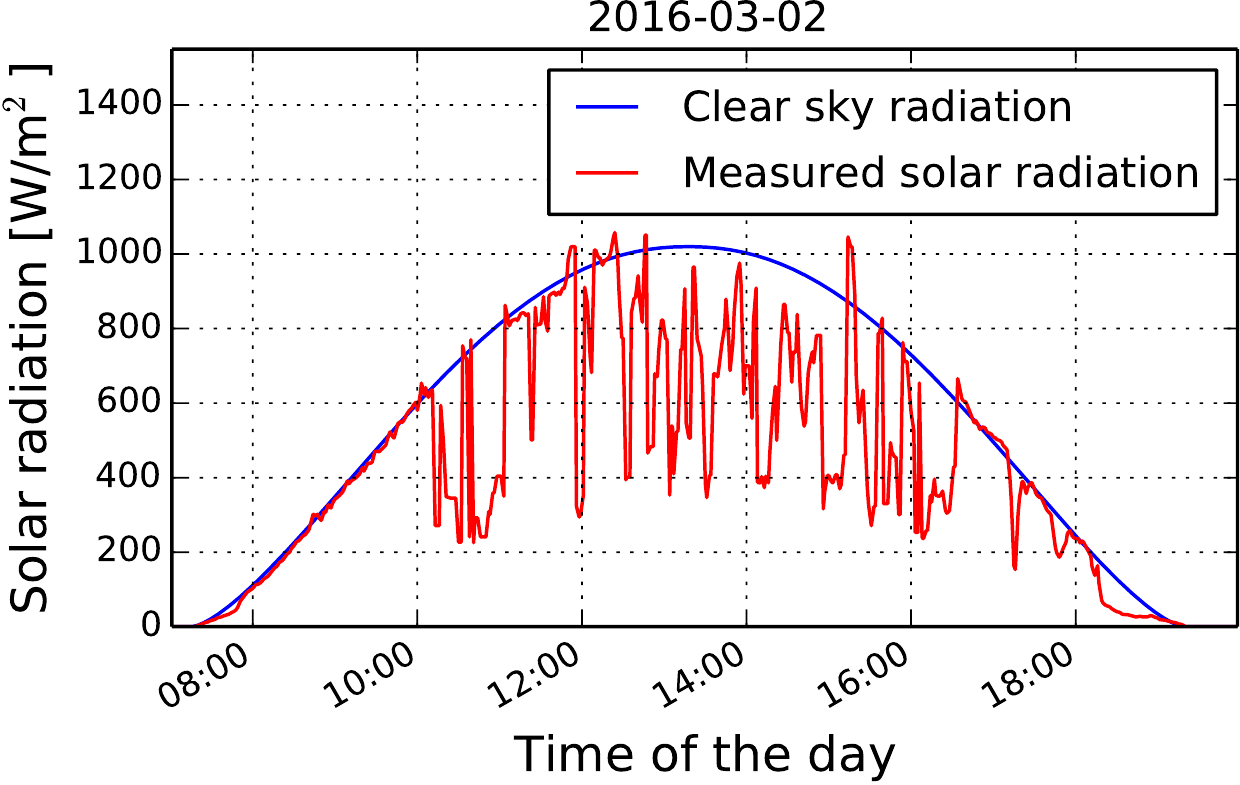}
\caption{Clear-sky model and measured solar radiation as on 02-Mar-2016. The deviation of the measured radiation from clear-sky model is because of the occluding cloud coverage.}\label{fig:oneDay}
\end{center}
\end{figure}

The Cloud Radiative Effect (CRE) is defined as the difference between the measured- and clear-sky solar irradiance. The measured solar radiation values are obtained from the weather station located at the rooftop of Nanyang Technological university building ($1.3455^{\circ}$N $103.6797^{\circ}$E). Larger CRE leading to a decrease in the amount of generated solar energy indicates the presence of occluding clouds. Figure~\ref{fig:oneDay} shows the measured- and clear-sky solar irradiance on 02-Mar-2016. We calculate the deviation of the measured solar radiation from the clear sky model, and use it for our subsequent analysis.

\begin{figure*}[htb]
\begin{center}
\noindent
\includegraphics[height=4.4cm]{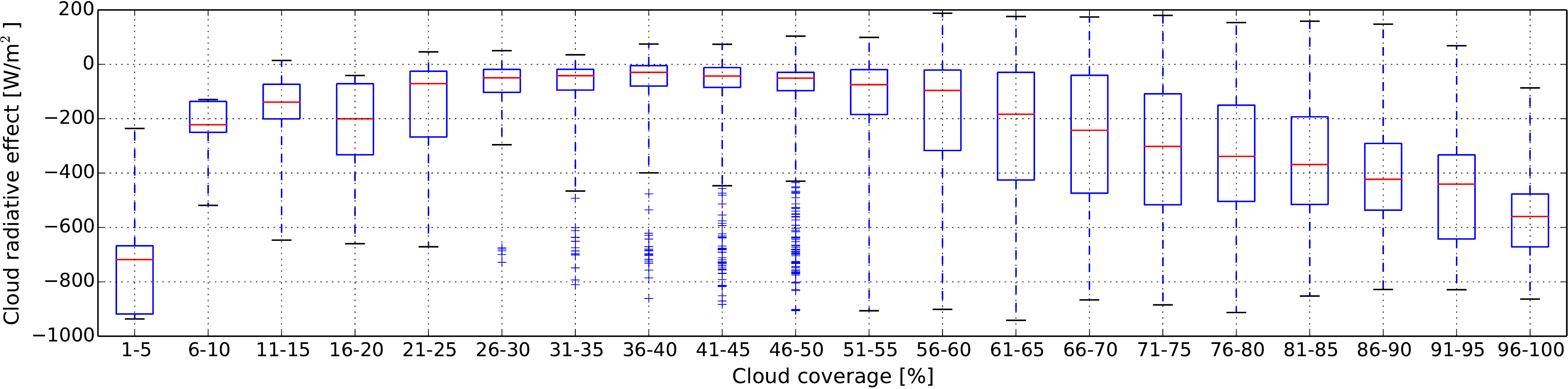}
\caption{Relation between cloud radiative effect and corresponding cloud coverage obtained from sky camera. The lower and upper edge of each box indicate \nth{25} and \nth{75} percentile of the data, and the central mark is the median. We observe a higher deviation in CRE value, with an increase in the cloud coverage.\label{fig:boxPlot}}
\end{center}
\vspace{-0.6cm}
\end{figure*}

\subsection{Computation of Cloud Coverage}
\begin{figure}[htb]
\begin{center}
\noindent
\includegraphics[width=1.7in]{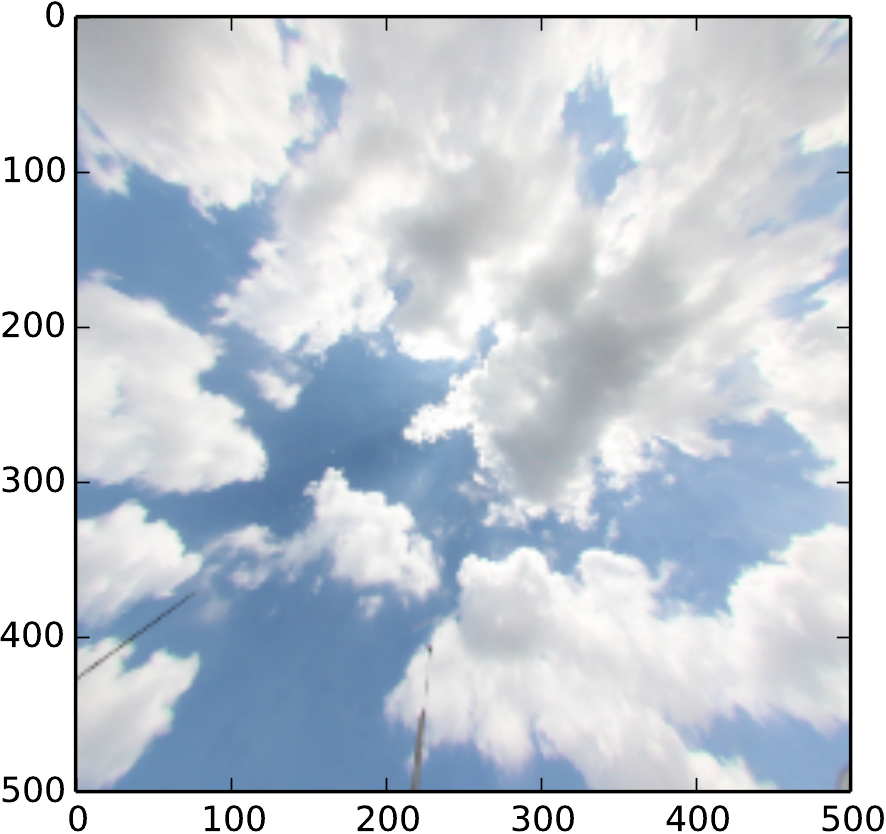}
\includegraphics[width=1.7in]{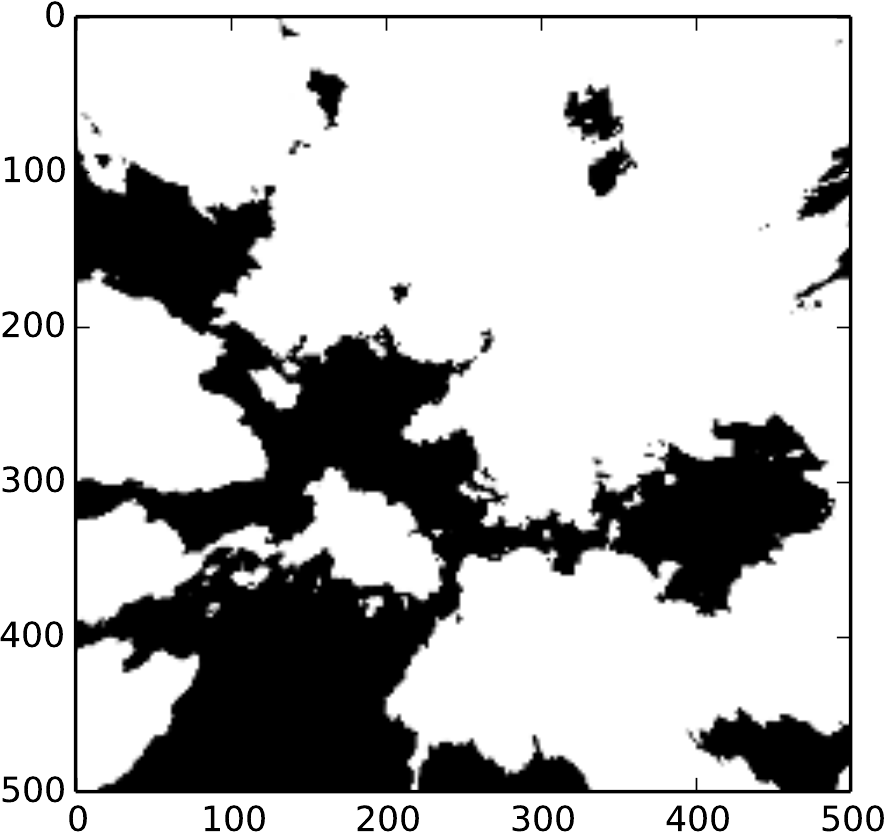}
\caption{Fish-eye lens image is undistorted to a $500 \times 500$ image using camera calibration function. This undistorted image is converted to the binary image using $(B-R)/(B+R)$ channel, constituting a cloud coverage of $71.1$\%.}\label{fig:cloudCover}
\end{center}
\vspace{-0.3cm}
\end{figure}

In most cases, the cloud coverage is reported manually through human observations. It is calculated in terms of oktas~\footnote{Okta is a unit of measurement for cloud coverage; where $0$ indicates completely clear-sky and $8$ indicates overcast condition.}, and is reported in the regular meteorological reports. However, we use ground-based sky cameras for automatic cloud coverage computation. The images obtained from our collocated sky camera~\cite{IGARSS2015a} are generally distorted owing to the wide-angle lens. Therefore, we undistort the captured image using its camera calibration function. Post undistortion, we use the ratio of $(B-R)/(B+R)$ for detecting cloud pixels, where $B$ and $R$ are the blue and red color channels respectively of the undistorted image. We perform clustering on this ratio image to generate the binary map~\cite{JSTARS2017}, as shown in Fig.~\ref{fig:cloudCover}. Finally, we calculate the cloud coverage percentage as the percentage of cloud pixels in the generated binary map. 

\section{Relation between Cloud Coverage and Radiative Effect}
In this section, we will statistically analyze the correlation between the cloud coverage and its corresponding CRE value. We calculate the CRE values as described in Section~\ref{sec:CREffect}, and the corresponding cloud coverage percentages from the sky camera images. We perform our experiments~\footnote{~The source code of all simulations in this paper is available online at \url{https://github.com/Soumyabrata/cloud-radiative-effect}.} by considering all the days in the month of March 2016, and selecting the images from the time period $08$:$00$ AM to $06$:$00$ PM. We consider a total of $6365$ images for our analysis. 

Figure~\ref{fig:boxPlot} shows the boxplot between cloud coverage and the corresponding CRE values. From the results, we can clearly observe the general trend of CRE as compared to cloud coverage. We observe that the magnitude of median CRE value gradually increases with increasing cloud coverage. This explains that, in the instant of overcast sky conditions, there is a huge reduction in the obtained solar radiation as compared to the clear-sky solar radiation. Similarly, the CRE is less for low cloud coverage scenarios. 

The variation observed in the individual boxplot is the effect of the time of a day. When similar cloud coverage is present at two different times of the day, the difference between observed solar radiation and clear sky values will vary accordingly. It will be high during mid-day as compared to the morning- and evening- hours. This is because the clear sky model has a cosine response with respect to time of the day. We also note that there is a high CRE value for the first box of the boxplot (constitutes only $0.1$\% of the entire data); this is owing to possible mis-classifications in cloud detection of the undistorted images.

\section{Conclusion}
In this paper, the relationship between cloud coverage and the cloud radiative effect is investigated. We obtain the cloud coverage percentage automatically from the ground-based sky camera. The CRE values are calculated from the measured solar radiation and clear-sky model. Extensive results show that higher cloud coverage leads to larger cloud radiative effect. Future works involve modeling this relationship, by considering more statistical data into consideration, and by studying the effect of seasonal- and time-of-day variations.

\bibliographystyle{IEEEbib}
\end{document}